# Internet Attacks:
# A Policy Framework for Rules of Engagement[*]

William Yurcik[1]   David Doss[2]

Department of Applied Computer Science
Illinois State University
{wjyurci,dldoss}@ilstu.edu

ABSTRACT:

Information technology is redefining national security and the use of force by state and nonstate actors. The use of force over the Internet warrants analysis given recent terrorist attacks. At the same time that information technology empowers states and their commercial enterprises, information technology makes infrastructures supported by computer systems increasingly accessible, interdependent, and more vulnerable to malicious attack. The Computer Security Institute and the FBI jointly estimate that financial losses attributed to malicious attack amounted to $378 million in 2000. International Law clearly permits a state to respond in self-defense when attacked by another state through the Internet, however, such attacks may not always rise to the scope, duration, and intensity threshold of an armed attack that may justify a use of force in self-defense.

This paper presents a policy framework to analyze the rules of engagement for Internet attacks. We describe the state of Internet security, incentives for asymmetric warfare, and the development of international law for conflict management and armed conflict. We focus on options for future rules of engagement specific to Information Warfare.

We conclude with four policy recommendations for Internet attack rules of engagement: (1) the U.S. should pursue international definitions of "force" and "armed attack" in the Information Warfare context; (2) the U.S. should pursue international cooperation for the joint investigation and prosecution of Internet attacks; (3) the U.S. must balance offensive opportunities against defensive vulnerabilities; and (4) the U.S. should prepare strategic plans now rather than making policy decisions in real-time during an Internet attack.

---

[*] supported in part by grants from DARPA # F30602-97-1-0257, NASA #NGT-30019, State Farm Insurance, and the John Deere Corporation

[1] Assistant Professor and corresponding author; additional contact information: telephone/fax: 309-556-3064/3864; postal mail: 45 Oak Park Road, Bloomington IL 61701 USA
[2] Associate Department Chair and Associate Professor, He is also a retired Lt. Commander US Navy (SSN).

1.0    Introduction

The development of "information warfare" (IW) presents policy issues related to a nation's efforts to both execute and respond to certain IW attacks, specifically those attacks using computers, telecommunications, or networks to attack adversary information systems.  The development of information technology makes it possible for adversaries to attack each other in new ways and with new forms of damage and may create new targets for attack.  Furthermore, the ability of signals to travel across international boundaries and affect systems remotely makes the application of international law problematic since the longstanding principle of national territorial sovereignty is difficult to apply to the Internet.

We define IW as an attack on information systems for military advantage using tactics of destruction, denial, exploitation, and/or deception.[YURCIK97]  An information system is defined as the life cycle of information generation/utilization based on a goal-directed activity.  Fischer (1984) defines the life cycle of information in eleven distinct stages:

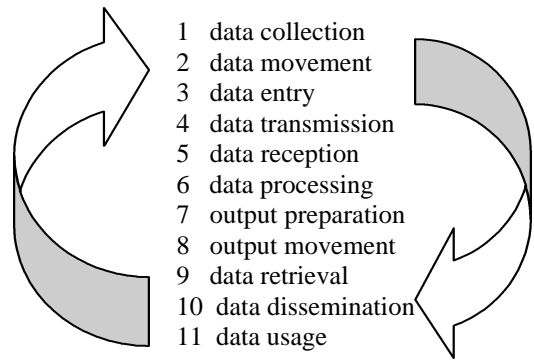

1   data collection
2   data movement
3   data entry
4   data transmission
5   data reception
6   data processing
7   output preparation
8   output movement
9   data retrieval
10  data dissemination
11  data usage

The information cycle is vulnerable to an IW attack at each and every stage.  While current research is searching for robust protection solutions for each stage in the information cycle, there is a systemic problem in that for every new protection a new threat can be developed in response.

The US is potentially vulnerable to IW attack because it is more dependent on information systems than any other country in the world.   Industrial society is increasingly dependent upon open large-scale infrastructure systems that operate interdependently such that a failure in one infrastructure can crossover into another infrastructure (i.e., a power grid failure affecting transportation systems).  Although the efficiency and quality of services these infrastructures provide is increasing, the interconnectivity of computer systems also increases accessibility to IW attack.

A Presidential Commission was formed on July 15, 1996 by Executive Order 13010 to issue recommendations on how to best protect from IW attack critical national infrastructures considered so vital that their incapacity or destruction would have a debilitating impact on the defense and economic security of the U.S.  In response to this commission's report, Presidential Decision Directive 63 (PDD63) establishes new governmental structures including:

> NIPC – National Infrastructure Protection Center
> Focal point for threat assessment, warning, investigation, and response to threats or attacks against critical infrastructures
>
> CIAO – Critical Infrastructure Assurance Office
> Responsible for integrating the various infrastructure sector plans into a National Infrastructure Assurance Plan and coordinating analyses of the U.S. government's dependencies on critical infrastructure

In addition, the U.S. Department of Defense (DoD) has budgeted billions of dollars to IW and all the military services have formed distinct IW organizations which are drafting IW military strategies.[YURCIK97] Specifically, the Defense Advanced Research Projects Agency (DARPA) funds on-going national research in defensive information survivability where survivability is defined as the ability of a system to continue functioning when part of system is unavailable due to failure, accident, or malicious attack.

International concern has also been raised over the potential for damage and disruption caused by military operations against and through information systems linked by the Internet. The following is excerpted from a United Nations General Assembly resolution that resulted from an initiative by the Russian Federation. This resolution calls upon Member States to promote multilateral "consideration of existing and potential threats in the field of information security". Although couched in the terms of the use of force by non-state actors, it was initiated in the First Committee, which concerns itself with disarmament between Member States. The momentum of the international community will likely yield an attempt to prohibit certain military applications of the Internet but will more likely result in an informal yet better understanding of the application of international law to the use of force via the Internet by Member States.

> Developments in Telecommunication and Information in
> the Context of International Security[3]
>
> Noting that considerable progress has been achieved in developing and applying the latest information technologies and means of telecommunication…
>
> Expressing concern that these technologies and means can potentially be used for purposes that are inconsistent with the objectives of maintaining international stability and security, and may adversely affect the security of States…

---

[3] General Assembly Resolution 5/70, United Nations 53$^{rd}$ Session, UN document A/RES/53/70 (1998). This resolution was an initiative of the Russian Federation that originated in the First Committee (Disarmament and International Security) of the UN General Assembly.

> Considering that it is necessary to prevent the misuse or exploitation of information resources or technologies for criminal or terrorist purposes…
>
> 1. Calls upon Member States to promote at multilateral levels the consideration of existing and potential threats in the field of information security;
>
> 2. Invites all Member States to inform the Secretary-General of their views and assessments of the following questions:
>
> (a) General appreciation of the issues of information security;
>
> (b) Definition of basic notions related to information security, including unauthorized interference with or misuse of information and telecommunications systems and information resources;
>
> (c) Advisability of developing international principles that would enhance the security of global information and telecommunications systems and help combat information terrorism and criminality.

While the number of Internet attacks being investigated by the U.S. Federal Bureau of Investigation (FBI) doubled from 1998 to 1999 (from 547 to 1,154), there has not been a consensus that this increase represents a aggregate threat to critical infrastructures.[FBI] Based on a March 2001 survey of 538 major organizations, the Computer Security Institute (CSI) with the assistance of the FBI estimated that the total losses attributable to computer crime was $378 million for calendar year 2000.[CSI01] Public awareness was achieved in February 2000, when a series of coordinated denial-of-service IW attacks were launched against major US corporations.[4] Not only did the attacks prevent 5 of the 10 most popular Internet websites from serving its customers but the attacks also slowed down the entire Internet - Keynote Systems measured a 60% degradation in the performance of the 40 other websites that had not been attacked.[NELSON00] While the consensus analysis of these IW attacks is that they were technologically unsophisticated (executing a downloadable program), it is particularly disturbing the ease at which major corporate operations can be disrupted and the lack of defenses to prevent such attacks from re-occurring in the future. These attacks made newspaper headlines and lead to a White House meeting with leading E-Commerce parties.

The remainder of this paper is organized as follows: Section 2 proposes reasons for the poor state of security on the Internet, an environment ripe for IW. Section 3 introduces "asymmetric warfare" as motivation and gives examples of significant events that have already occurred. Section 4 cites international law relevant to the "use of force" and its application to IW. Section 5 discusses counter-offensive IW strategic options with their relevant ramifications. We close with a summary of conclusions and recommendations in Section 6.

---

[4] The companies in the order they were attacked are: Yahoo! (2/7/00), eBay (2/8/00), Buy.com (2/8/00), Amazon.com (2/8/00), CNN (2/8/00), ZDNet.com (2/9/00), E*Trade (2/9/00), Excite At Home (2/9/00), and Datek (2/9/00).

## 2.0 What Is Going Wrong With Security on the Internet?

If history is any indication, the information technology community is incapable of constructing networked information systems that can consistently prevent successful attacks. Even an organization where network and computer security is paramount such as the US DoD has continuously demonstrated how susceptible it is to attack.[5] There are many factors which contribute to the poor state of security on the Internet, but here we focus on only the top four factors.

First, most software today is tested for security (assurance testing) by the penetrate-and-patch approach – when someone finds an exploitable security "hole" the software manufacturer issues a patch.[REITER99] The intellectual complexity associated with software design, coding, and testing virtually ensures the presence of "bugs" in software that can be exploited by attackers. This approach has proved inadequate since after-the-fact security leaves bug vulnerabilities open until they are exploited. However, software manufacturers find this approach economically attractive – why invest time and money in assurance testing if consumers are not willing to pay a premium.

Second, software vulnerabilities will be discovered in commercial-off-the-shelf (COTS) and public-domain products whose internal structures are widely available and hence can be analyzed by attackers. The ability to discover software vulnerabilities and the homogeneity of software from the same manufacturer (i.e., Microsoft, Sun) make it possible for a single-attack strategy to have a wide-ranging and devastating impact.

Third, poor system administration practices result in a system remaining susceptible to vulnerabilities even after corresponding patches have been issued from software manufacturers. An otherwise secure system that does not fix a previously unknown vulnerability after a publicly announced patch has been released from the software manufacturer is instantly an open target for exploitation. Poor maintenance practices can neutralize existing security safeguards.

Lastly, even if special software from one manufacturer is "assured", it is executed on the same system with "non-assured" software such that new vulnerabilities are introduced into the system as a whole.[6][REITER99] When COTS products are incorporated as components of a larger system, the entire system become vulnerable to attacks based on the exploitable COTS bugs again making it possible for a single-attack strategy to have a wide-ranging and devastating impact.

---

[5] Specifically network security refers to protection of traffic as it traverses a network end-to-end or between nodes and computer security refers to operating system authentication of users (identity) and file system privileges. However, since networks consist of special-purpose computers (switches, routers, hubs) and computers rely on network/bus connectivity for peripherals and distributed services (Email, www, file system) we will not continue to make a distinction between network and computer security for the purposes of this paper. Later in the paper we discuss multiple examples of successful attacks against the US DoD.

[6] Developing systems via software integration and reuse rather than customized design and coding is a cornerstone of modern software engineering.(Software Engineering Institute at Carnegie Mellon University)

The natural escalation of offensive threats versus defensive countermeasures has demonstrated time and again that no practical systems can be built that are invulnerable to attack. Despite industry's efforts (or in spite of them), there can be no assurance that any system will not be compromised. Thus, the traditional view of information systems security must be expanded to encompass the specification and design of survivable systems that can survive attacks that cannot be completely repelled. The ubiquity of the Internet has only acerbated this problem by enabling automated, anonymous, and remote attacks by hostile nations who previously did not have a non-conventional military option.

3.0     Asymmetrical Warfare

In 1974 the DoD's Advanced Research Projects Agency (ARPA) and a number of universities began building the ARPANET, defining and standardizing the TCP/IP protocols and writing Requests for Comments (RFC) standards. By 1985, a fully functional Internet existed and was being used by academics and the military. Between 1986-1990, the U.S. Army installed TCP/IP on its IBM mainframes at its finance centers, logistic supply centers, and central data processing centers which up to this point had used only IBM's proprietary protocols (SNA and Bisync). Within six weeks of the Iraqi invasion of Kuwait in 1991, the U.S. Army Information Systems Command (USAISC) had designed and made fully operational a TCP/IP network in Saudi Arabia using COTS products. This network was a significant complement to existing Command-and-Control systems and satisfied battlefield requirements for unclassified logistics data as well as being reconfigurable in "real-time" as the situation changed. Just as the Patriot missile became the hero of the U.S. Army's Desert Storm Weapons systems, the Internet had become the hero of Operation Desert Storm's information systems. [WEISSERT92] Further military development of the Internet beyond Operation Desert Storm is discussed throughout this paper.

However, there are significant strategic problems with military dependence upon the Internet. The open architecture of the Internet is ideally suited for "asymmetrical warfare", a military term referring to the indirect use of force on an opponent as opposed to a direct force-on-force confrontation. The conventional military force (Army, Navy, Air Force, Marines) of the U.S., as exhibited in recent operations, is unquestionably supreme as the world's remaining superpower.[7] However, much of the U.S. military capability is supported by underlying information systems with their corresponding vulnerability to attack. Information systems are so critical to military operations that it is often more effective to attack an opponent's information system than to concentrate on directly destroying conventional military forces. International law has long recognized the control and use of information such as signal intelligence, communications intelligence, electronics intelligence, foreign instrumentation signals intelligence, and imagery intelligence.[8] There is a perception within military circles that control and use of information may be more important than air superiority in previous wars.[YURCIK97]

---

[7] Even with drastic cuts the U.S. military budget is larger than all other nation's military budgets combined.

The tactical and strategic use of this intelligence gathered via networked information is embraced in the operational DoD concept of "information operations". Over the Internet, attackers can anonymously collect a wide range of intelligence (sensitive/classified information), manipulate data to deceive decision-makers, influence public opinion, and even cause physical damage. "Tapping" into an adversary's command-and-control system provides the ultimate information about enemy battle plans and has been the motivation behind trillions of dollars of U.S. spy operations during the Cold War.[9] The longer the "tap" remains undetected the longer the intelligence flows. Psychological operations (PSYOPS) include implanting false information about troop locations and battlefield conditions into enemy computers to subvert adversary decision-making.

An attacking state with a capability in information technology does not need conventional military force to execute an organized, large-scale attack on U.S. military, commercial, or civilian targets from remote locations abroad. The asymmetry of IW is highlighted in that an attacker does not need an underlying information infrastructure that would itself be vulnerable to counter-attack. With an investment of a few thousand dollars in consumer personal computers, a dedicated and persistent hostile attacker can gain access to almost any Internet-linked information infrastructure of any state in the world and yet lack sufficient targets for any IW retaliation.[SHARP99]

The Director of the U.S. Central Intelligence Agency (CIA), George Tenet, has repeatedly stated in testimony before Congressional hearings that more than a dozen countries – among them Russia, China, Iraq, Iran, Cuba, France, and Israel – are developing significant IW capabilities. A senior CIA official cited a Russian General who compared the disruptive effects of IW on transportation and the electric grid as comparable to the effects of a nuclear weapon.[STROBEL00] China is considering the creation of a fourth branch of its armed service devoted to IW.[STROBEL00] The following is quoted from congressional testimony:

---

[8] Signal intelligence (SIGINT) is intelligence information comprising either individually or in combination all communications intelligence, electronics intelligence, and foreign instrumentation signals intelligence, however transmitted. Communications intelligence (COMINT) is technical and intelligence information derived from foreign communications by other than the intended recipients. (ELINT) is technical and intelligence information derived from foreign non-communications electromagnetic radiation emanating from other than nuclear detonation or radioactive sources. Foreign instrumentation signals intelligence (FISINT) is technical and intelligence information derived from the intercept of foreign electromagnetic emissions associated with the testing and operational deployment of non-US aerospace, surface, and subsurface systems such as telemetry, beaconry, electronic interrogators, and video data links. Imagery intelligence (IMINT) is intelligence derived from the exploitation of collection by visual photography, infrared sensors, lasers, electro-optics, and radar sensors such as synthetic aperture radar wherein images of objects are reproduced optically or electronically on film, electronic display devices, or other media. Definitions from *National Security Law* by John Norton Moore et. al., editors 1990.

[9] For one source of information see *Blind Man's Bluff* by Sherry Sontag and Christopher Drew (HarperCollins 1998) which documents Cold War operations by US spy submarines including "taps" on Russian Command and Control undersea cables.

> "We are detecting, with increasing frequency, the appearance of doctrine and dedicated offensive cyberwarfare programs in other countries. We have identified several (countries), based on all-source intelligence information, that are pursuing government-sponsored offensive cyberprograms… Information Warfare is becoming a strategic alternative for countries that realize that, in conventional military confrontation with the United States, they will not prevail. These countries perceive that cyberattacks launched within or outside of the U.S. represent the kind of asymmetric option they will need to level the playing field during an armed crisis against the U.S.. … The very same means that the cybervandals used a few weeks ago could also be used on a much more massive scale at the nation-state level to generate truly damaging interruptions to the national economy and infrastructure."[10]

Examples of IW involving the U.S. include the following: During Operation Desert Storm (between April 1990 and May 1991), hackers from the Netherlands penetrated 34 U.S. military sites including military supply systems[11] and gained such information as the exact location of U.S. troops, their weapons, and the movement of U.S. warships.[DENNING99] During the military exercise Eligible Receiver in June 1997, the NSA demonstrated that a hostile enemy state could disrupt computer operations at major military commands, cause large-scale blackouts, and interrupt emergency service in Washington D.C. in several other cities in the U.S.[GRAHAM98,MCCULLOGH00] In February 1998 two intruders accessed (but did not compromise) sensitive DoD systems during planning for military airstrikes in Iraq. [GRAHAM98]

The first reported organized attack over the Internet by one state against computers in another state was a highly organized attack by Indonesia in January 1999 against the virtual country domain of East Timor supported by non-government computers in Ireland.[SHARP99] Burma's military junta is blamed for the "Happy 99" Email virus sent to dissidents.[STROBEL00] In the summer of 1999, after Taiwan's President Lee Teng-Hui voiced support for the islands independence, China began to launch IW attacks aimed at altering official Taiwanese government websites which culminated in a national blackout and crashing of financial networks.[STANTON00] In January 2000, Azerbaijan hackers tampered with dozens of Armenian-related websites including host computers in the U.S.[12][STROBEL00] Internet sites belonging to the Falun Gong spiritual movement based in Long Island New York have been attacked by the People's Republic of China's Secret Police in Beijing.[13]

---

[10] John Serabian, the CIA's information operations issue manager, in testimony before the Joint Economic Committee of Congress 3/4/00.
[11] US military forces are particularly dependent upon non-military systems for deployment and logistics.
[12] Azerbaijan and Armenia fought a war over the disputed territory of Nagorno-Karabakh. Jonathon Peizer of the Open Society Institute whose office was affected by the Azerbaijan IW attack stated, "(It is) the first precedent of a physical battle going online."[STROBEL00]
[13] These attacks were traced back to the XinAn Information Service Center which is part of the Ministry of Public Security – China's Secret Police.[STROBEL00]

## 4.0 What Constitutes the Use of Force on the Internet?

When are states allowed to use force against each other on the Internet? For example, the United States has the right under existing international law to respond with military force if a state destroyed or significantly damaged the New York Stock Exchange by the use of computer viruses or other network attack. Although the argument that IW would violate neutrality is strong, encroachments beyond a nation's borders that violate neutrality have, in the past, been physical intrusions. For example, orbital remote sensing, which may include the bombardment of a country's territory with radar or other forms of electromagnetic radiation, is permissible during war or peace.

### 4.1 International Law of Conflict Management

The law of conflict management, jus ad bellum, is a set of rules that govern the resort to armed conflict and determine whether the conflict is lawful or unlawful in its inception. [YURCIK97, SHARP99] Articles 2(4) and 51 of the Charter of the United Nations codifies contemporary jus ad bellum. If a state activity is a use of force within the meaning of Article 2(4) of the Charter, it is unlawful unless it is an exercise of that state's right of self-defense or unless specifically authorized by the Security Council under its Chapter VII authority.

Article 39 of the Charter of the United Nations empowers the UN Security Council to authorize the use of force to maintain international peace and security under international law at a threshold lower than when states are authorized to use force in self-defense. A blockade, for example, that would be otherwise unlawful under international law if committed by a state unilaterally would be lawful if authorized by the UN Security Council under its Chapter VII authority. A "virtual" blockade became an Internet issue in 1999 when Loral Space & Communications and UUNET were involved in negotiations with the U.S. State and Treasury Departments to disconnect Yugoslavia from the Internet.[ZERO99]

Customary international law requires that all uses of force be "necessary" and "proportional" and it prohibits the use of force for retaliatory or punitive actions.[YURCIK97] The principle of necessity requires that the use of force for self-defense is justified by a continuing threat which has not been alleviated with all other possible means. The principle of proportionality requires that the use of force be of the same intensity and magnitude necessary to promptly secure the limited military objective of self-defense given the possibility of collateral civilian destruction. It is emphatically not a requirement of the principle of proportionality that any "necessary" self-defense limit itself to weapon systems employed by the aggressor or matching troop strength levels.[SHARP99] The principle of proportionality balances the need to attack a "military objective" with the collateral effects of civilian deaths and property damage. In the absence of civilians, the principle of proportionality imposes no limitations on the use of force. A lawful "military objective" includes combatants and:

> those objects which, by their nature, location, purpose, or use, effectively contribute to the enemy's war-fighting or war-sustaining capability and whose total or partial destruction, capture, or neutralization would constitute a definite military advantage to the attacker under the circumstances at the time of the attack. [SHARP99]

While the principle of proportionality limits the use of force given civilian collateral damage, it does not prohibit any damage to civilians including death, injury, or property. While civilians and civilian property may not be the objective of an attack, states may use force against civilians, civilian property, and activities that support or maintain an enemy state's warfighting capability. For example, states may use force during armed conflict against economic targets such as telecommunications, rail yards, bridges, ports, industrial plants producing warfighting products, and the electric power grid. It is commonly quoted that in the U.S. 95% of all military communications is carried over civilian infrastructure thus intermingling military/civilian targets. In most modern societies, much of the civilian infrastructure is used for military purposes and is thus subject to lawful attack during armed conflict if there is a military advantage to be gained by such an attack.

To the extent private industry and civilians and especially communications have supported war efforts, they have always been a lawful target under the law of armed conflict. Undersea cables, including those connecting belligerents with neutrals, have been interfered with during all naval wars since the Spanish American War.[GREENBERG98] For example, as World War I began in August 1914, the British cableship Telconia cut Germany's undersea cables and reeled in the loose ends to prevent repair. The infamous Zimmerman telegram intercepted from an undersea cable from Germany to Mexico was one of the precipitating events to bring the U.S. into World War I. However, the current difference is that modern society's increasing dependence on underlying computerized infrastructures makes more civilian infrastructures highly vulnerable and subject to justifiable attacks from remote locations via the Internet.

4.2     International Law of Armed Conflict

Any attack involving networks and telecommunications may implicate the International Telecommunication Union (ITU) and its underlying charter, the International Telecommunication Convention (ITC) which apply to international wire and radio frequency communications. The relevant jurisdiction of the ITU is spectrum interference – the ITU allocates frequency spectrum to prevent interference and states may not broadcast false or misleading signals. In practice, governments have conducted radio jamming in both peace and war for over 60 years beginning with Austria's efforts to block broadcasts from Nazi Germany in 1934.

The law of war, jus in bello, also commonly referred to as the law of armed conflict, is derived in part from Christian "just war" doctrine and governs the actual conduct of hostilities and has developed as customary international law over thousands of years.[SHARP99,YURCIK97]   The law of armed conflict is codified in two regimes: (1) the Hague Regulations that govern the means and methods of warfare, and (2) the Geneva Conventions that govern the protection of victims of war. The continued existence of war atrocities is not a failure of the rule of law but rather a failure of the international community to enforce the law.[SHARP99]

According to Common Article 2 of the four Geneva Conventions of 1949, an international armed conflict exists upon declaration of war, the occurrence of "any other armed conflict" between two or more parties (even if the state of war has not been declared), and in all cases of partial or total occupation even if met with no armed resistance. Although the terms 'war' and 'armed conflict' are frequently used interchangeably and refer to a state of hostilities that invokes the law of armed conflict, war refers to a state of de jure hostilities invoked by formal declaration by one party that creates an international armed conflict as a matter of law. In contrast, 'any other armed conflict' refers to a state of de facto hostilities invoked by the use of force by one party without formal declaration of war.

4.3     Evolution of U.S. Rules of Engagement

The consensus of the international community is that de facto armed conflict exists between states when the scope, duration, and intensity of force between them reaches the level of armed attack within the meaning of Article 51 of the UN Charter. What constitutes a use of force of a scope, duration, and intensity that constitutes an armed attack and triggers the law of armed conflict is a question of fact that must be subjectively analyzed in each and every case in the context of all relevant law and circumstances.

In 1983, the U.S. Department of State defined 'armed conflict' as "any situation in which there is hostile action between the armed forces of two parties, regardless of duration, intensity, or scope of the fighting and irrespective of whether a state of war exists between the two parties."[14]

In 1997, the U.S. was adopted the following position during the ratification process of the Chemical Weapons Convention:

> "In accordance with Condition (26) on Riot Control Agents, I have certified that the United States is not restricted by the Convention in its use of riot control agents in various peacetime and peacekeeping operations. These are situations in which the United States is not engaged in a use of force of a scope, duration, and intensity that would trigger the laws of war with respect to U.S. forces."[15]

This reversal of position from 1983 to 1997 was a result of the increasing use of U.S. troops in peace-keeping efforts world-wide. The intent was to allow U.S. peace-keeping personnel to use force in self-defense. Otherwise, U.S. peacekeeping personnel would become lawful targets of a "de facto armed conflict".

---

[14] This quote is taken from a press release from the Department of State in the case of U.S. Navy Lieutenant Robert O. Goodman Jr. who was shot down and held by Syria. The issue of Goodman's status as a 'prisoner of war' were raised under the Geneva Convention.

[15] President's Certifications and Report to the Congress in Connection with the U.S. Senate Resolution of Ratification of the Chemical Weapons Convention, April 25, 1997.

4.4     A Special Case: Perfidy

A special case of manipulating information is perfidy. Although diversionary tactics are unquestionably permissible in war, not all acts of deception are. Perfidious acts include feigning protected status for purposes of attacking the enemy.[SHARP99]   Certain IW attacks may fall under this category.  For example, manipulating enemy information systems into identifying U.S. military targets as neutral would be perfidy.[SHARP99]

5.0     Current Rules of Engagement

There are no characterized rules of engagement for IW conflicts which can take many different forms (isolated operations or undeclared wars between states).  We next examine more closely the current general rules of engagement for the U.S. military and then examine options for future rules of engagement specific to IW.

"Hostile act" and "hostile intent" are the operational concepts integral to the ROE of a state that acts upon its inherent right of self-defense.  The "Standing Rules of Engagement for U.S. Forces" (SROE) define hostile act and hostile intent as follows:

> Hostile Act: An attack or other use of force against the United States,
> ,US forces, and, in certain circumstances, US nationals, their property,
> US commercial assets, and/or other designated non-US forces, foreign
> nationals and their property.  It is also force used directly to preclude or
> impede the mission and/or duties of US forces,  including the recovery
> of US personnel and votal US Government property.[SROE00]

Examples of hostile acts include blockades; the destruction of early warning or command-and control systems; and any use of military force. Although all uses of force are not of the scope, duration, and intensity to constitute an armed attack, state activities that constitute a hostile act are a use of force that constitutes an armed attack and a state which is the object of that hostile act has the right to use necessary and proportional force to respond in self-defense.[SHARP99]

> Hostile Intent:  The threat of imminent use of force against the
> United States, US forces, and in certain circumstances,  US nationals,
> their property, US commercial assets, and/or other designated non-US
> forces, foreign nationals, and their property.  Also, the threat of force to
> preclude or impede the mission and/or duties of US forces, including
> the recovery of US personnel or vital USG property.[SROE00]

> Declaring Forces Hostile:  Once a force is declared hostile by
> appropriate authority, US units need not observe a hostile act or a
> demonstration of hostile intent before engaging that force.  The
> responsibility for exercising the right and obligation of national self-
> defense and as necessary declaring a force hostile is a matter of the

> utmost importance. All available intelligence, the status of international relationships, the requirements of international law, an appreciation of the political situation, and the potential consequences for the United States must be carefully weighed. The exercise of the right and obligation of national self-defense by competent authority is separate from and in no way limits the commander's right and obligation to exercise unit self-defense.[SROE00]

Isolated verbal threats and threats of the use of force that are conveyed by troop/weapon movements and alliances do not constitute hostile intent. Examples of hostile intent include the use of force conveyed by massing of troops at a border, weapon status/targeting indications, the tactical use of fire control radar, and interference with early warning/command-and-control systems. State activities that convey hostile intent constitute a threat to use force and a state, which is the object of that hostile intent, has the right to use necessary and proportional force to respond in anticipatory self-defense.

5.1     Future Rules of Engagement: Counter-Offensive IW

Eventually a nation state will determine that the potential gains of an IW attack on the U.S. (or those of our allies and/or neighbors) outweigh the potential risks of such actions. In a defensive stance, the U.S. can manage these risks – the risks of being identified and the risks of punishment. Identification is largely a technical problem at the current time so deterrence is the policy option we will focus upon.

The Pentagon's current "purely passive" policy of prohibiting the DoD from mounting a retaliatory counter IW attack could result in "severe consequences for U.S. military capabilities" according to a 1999 report issued by the National Research Council (NRC).[NRC99]   Under current policies DoD can only track IW attacks and when the attacker can be identified, the DoD must transfer the responsibility of prosecuting non-state actors to law enforcement officials.

The NRC report, which includes a review of DoD command, control, communications, computer, and information systems ($C^4I$) policies and practices added weight to statements by top Pentagon officials that attacks against DoD information systems that are critical to a new doctrine of information-based warfare continue to escalate while passive defensive measures lag. Simply protecting information systems from attack may not work in the future:

> "DoD is in an increasingly compromised position. The rate at which information systems are being relied on outstrips the rate at which they are being protected… The time needed to develop and deploy effective defenses in cyberspace is much longer than the time required to mount an attack" [BREWIN99]

Although the NRC report raises many policy issues, it specifically urges the Pentagon leadership to "review the legal limits in its ability to defend itself and its $C^4I$ infrastructure from attack. [Then DoD] should take the lead in advocating changes in national policy (including legislation if necessary) that amend the rules of engagement specifying the circumstances under which force [would be] an appropriate response to a cyber-attack."[16] [BREWIN99]

Maintaining a credible ability to use force on the Internet is lawful and a fundamentally important aspect of deterrence and international peace and security. Even clearly defined laws and norms of behavior are not adequate to control aggressive states, terrorists, and criminals. Accordingly, states must maintain a strong and credible ability to use force when necessary in self-defense to ensure the rule of international law is enforced. The strategic policy decision to threaten or use forces in not inherently unlawful or evil but rather essential to the maintenance of international peace and security. Indeed, the collective use of force by the international community is the core principle upon which the Charter of the United Nations is built. International law outlaws only the aggressive use of force and it specifically acknowledges the use of force in self-defense.

There is no consensus over the point in time at which such a counter-offensive IW attack in self-defense can be taken.[SHARP99] There are two options: (1) a preemptive attack before an anticipated attack occurs or (2) a second-strike attack in response after an attack has occurred or is in the progress of occurring. The right to respond in anticipatory self-defense or second-strike does not apply to the penetration of all government systems during peacetime but it should apply presumptively to those sensitive systems that are critical to a state's vital national interests. There is the most difficult tactical problem of not only identifying the attack and the attacker but also discerning intent since what may appear as a primary IW attack may in reality be pre-attack exploration, espionage, or vandalism.[17] The state-of-the-art in accurately predicting intent is based on studying patterns of IW attack. [GREENBERG99] presents a classification of IW attacks based on physical intrusiveness and physical destructiveness.

Some scholars conclude that the customary international law right of a preemptive self-defense is incorporated in Article 51 of the UN Charter while others conclude to the contrary. An example of a preemptive attack was the 1980 Israeli bombing of an Iraq nuclear power plant suspected of making weapon-grade plutonium. An example of a second-strike attack is the August 1998 U.S. bombing of terrorist camps in Afghanistan and Sudan in response to U.S. embassy bombings in Tanzania and Kenya. The U.S. justified these second-strike attacks as lawful under the right to self-defense as recognized in Article 51 of the UN Charter.

The U.S. Space Command in Colorado assumed the responsibility for offensive IW programs on October 1999. As part of the nation's national defense plan, the U.S. military is drafting plans to penetrate and disrupt computers of enemy nations.[MCCULLUGH00] The Pentagon's January 2000 announcement of its counter-offensive strategy had been quietly discussed for nearly a year and came at a time when military worries about hackers were at an all-time high due to anticipated terrorist attacks on Y2K (which never materialized).

---

[18] Retired Lt. General Carl O'Berry, former Air Force deputy chief of staff for C4I and a member of the NRC committee states, "We debated [the cyberattack recommendation]. We have foes with unlimited opportunities to attack while we maintain a defensive crouch [at a time] when some people think we are in a cyberwar."[BREWIN99]

[19] These issues lead John Pike, an analyst with the Federation of American Scientists (FAS) to state, "Does this mean the Pentagon will then start frying the home PCs of American teen-age hackers?"

Practically, a policy allowing the military to "hack-back" against an attacker has several problems: (1) the unanticipated, and possibly counter-productive, technical consequences of transmitting an attack against an attacker; (2) legal prohibitions against such action (deploying troops) on U.S. soil; (3) the need to get clearance from countries in which routers between the two principal parties are located; (4) technical difficulties in identifying the attacker; and (5) shortages of staff to do the work. Hack-back can take many forms such as transmitting a virus, denial-of-service or flooding attacks, chipping (hardware-based malicious software), and directed energy weapons that effectively destroy electronic systems. It is unexplored territory, no one knows how a nation would respond to an IW attack or counter-attack.

The U.S. military has signaled that its response to an IW attack would be devastating. The Pentagon's Office of General Counsel stated in the 1999 document, An Assessment of International Legal Issues in Information Operations, the consequences of a large-scale campaign of computer network attacks "might well justify a large-scale military response."[STANTON00]    However, the right of self-defense under customary international law may not always justify a large-scale military response due to the principles of necessity and proportionality. There are a spectrum of possible attack scenarios from threats of force to actual use of force and all levels in between. Figure 1 graphically depicts this spectrum of force/response scenarios adapted from [SHARP99]. For instance, a low-level verbal threat to use force, while not prohibiting an attack in self-defense, may not justify an armed response.

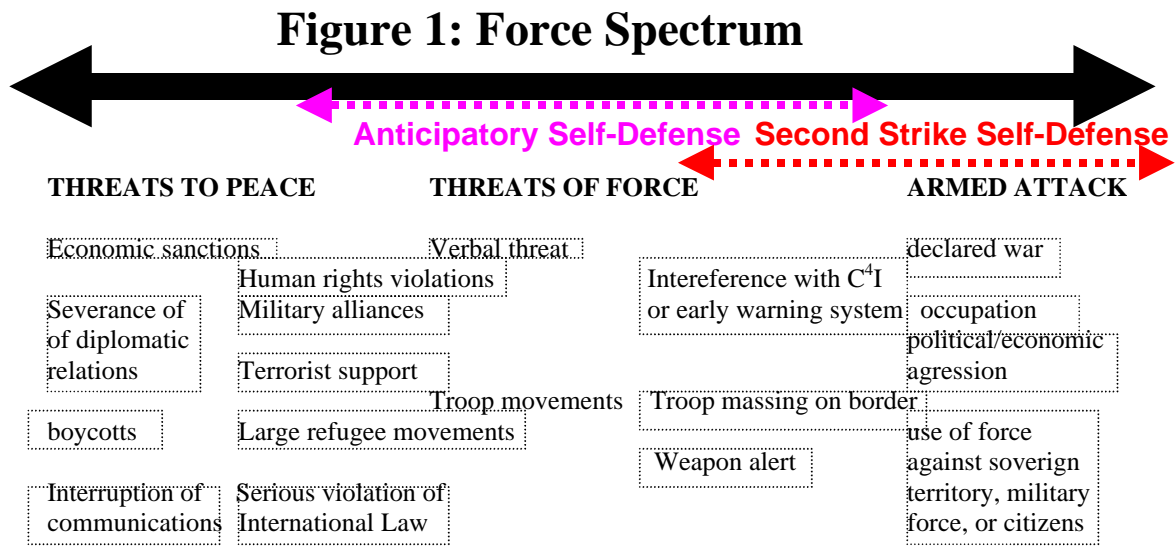

Figure 1: Force Spectrum

States will still need to explore offensive and counter-offensive capabilities if only to test their defensive measures adequately. Without knowing the extent of U.S. offensive capabilities or defensive vulnerabilities it is impossible to judge the desirability of constraining counter-offensive IW. If the U.S. leads in the technological development of offensive IW, as anecdotal evidence suggests, it would be counterproductive to draw attention to limiting IW weapons.[18] Other nations would be justifiably suspicious of the any U.S. effort to protect its technological advantage or restrain the development of their competing IW capabilities.

During the Kosovo bombing campaign last year, the Pentagon set up a high-level information operations cell for offensive IW attacks. According to Admiral James Ellis, NATO's Number 2 military commander during the war, "All the tools were in place." The official line is that the U.S. held back and by the time Pentagon lawyers had approved an IW attack against Serbia events had overtaken the need for such a strike.[STROBEL00] However, the details of part of the IW attack plan was leaked by senior intelligence officials to Newsweek.[VISTICA99] The attack plan called for U.S. government hackers to tap into Serbian Leader's Slobodan Milosevic's foreign bank accounts. The policy ramification of such an attack – a U.S.-sponsored plan to break into foreign banks – would be a breach of national sovereignty in friendly nations and opened the door to computer attacks on U.S. banks. Ultimately, the U.S. would have been the main loser if confidence in the world banking system were undermined by such attacks.

6.0     Summary

The terrorist attacks on the World Trade Center and Pentagon on September 11, 2001 changes our risk management profile for IW attacks. The details of this terrorist attack are just being investigated but it is clear that there were some IW components supporting this primarily physical kinetic attack. Before September 11, 2001, large scale distributed denial-of-service attacks on major corporate websites in February 2000 and continuous virus threats have been our experience with IW. These pure IW attacks did not cause critical or lasting damage, but they did take hypothesized threats out of the realm of the abstract and made them real. Given such attacks cannot currently be stopped and will likely increase in frequency, this paper has dealt with the defensive options available to responding to such attacks from non-state and state-sponsored attackers.

First, it has been established that IW attacks, even those attacks that are not directly lethal or physically destructive, constitute the use of force or armed attack in the United Nations Charter. Such attacks thus may be legal forms of coercion even in peacetime, and the use of conventional armed forces may not be an appropriate response to such attacks; indeed such a response might be considered a hostile act. It is equally unclear whether some of the damage that IW attacks could inflict, such as by disrupting government and private information systems, is the sort of damage that international humanitarian law is intended to restrain. A recommendation is that the U.S. could pursue international definitions of "force" and "armed attack" in the new IW context and create international protection schemes for critical information systems. Such international consensus will help establish when IW attacks can be conducted, what is off-limit for legitimate attacks, and how countries may respond to IW attacks. IW may be an appropriate area for arms control agreements in the future, however, the wide dissemination, commercial nature, and lack of verification mechanisms suggest that such a mechanism is premature at this time.

---

[20] Emmet Paige, Assistant Secretary of Defense for Command, Control, Communications, and Intelligence ($C^4I$) stated in 1995, "We [the US DoD] have an offensive [IW] capability, but we can't discuss it… [However] you'd feel good if you knew about it." (Munro, Neil. "*Pentagon Developing Cyberspace Weapons.*" Washington Technology, June 22, 1995.)

Second, since IW attacks are executed across international networks, the U.S. (among others) will need to rely upon foreign assistance to identify and respond to attackers. A recommendation is the U.S. should pursue international cooperation against IW attacks, encouraging joint investigation and prosecution of those state and non-state actors responsible for IW attacks, particularly terrorists and criminals.

Third, the ambiguous state of international law regarding IW may leave space for the U.S. to pursue offensive as well as counter-offensive IW. Conversely, it may permit adversaries to attack the U.S. and its systems. When considering policy options, U.S. decision-makers must balance offensive opportunities against defensive vulnerabilities, a balance that is beyond the scope of this paper. In this context, policy and law complement preparedness and ingenuity in the development of defensive IW technology, which will contribute to solving or worsening this problem.

Lastly, as in all forms of warfare, IW is dynamic. Since 1997 IW attacks have occurred with increasing frequency and responses to these attacks have evolved. It should be self-evident why the U.S. and the international community should be addressing the policy and technology issues of IW now rather than having to address them during an emergency where options and precedents are often set by exigencies rather than strategic planning.

7.0 Acknowledgments

We would like thank numerous members of the defense and intelligence communities who provided feedback on this paper but must remain anonymous. We would like to especially acknowledge the significant intellectual contributions of Walter Gary Sharp Sr. (the MITRE Corporation, Georgetown Law Center, and Ret. U.S. Marine Corps Judge Advocate who previously served as Deputy Legal counsel to the Chairman of the Joint Chiefs of Staff). Mr. Sharp has made the seminal contributions to understanding the use of force in the international legal context and has generously provided detailed feedback we have incorporated into this paper. Lastly we would like to thank Paul F. Capasso whose work at Harvard University in 1997 resulted in the first academic paper to specifically identify potential information warfare attacks on civilian infrastructure. Paul reviewed an early version of this paper in 2000. It should make U.S. citizens feel reassured that Paul is now working for the U.S. Space Command.